\UseRawInputEncoding
\documentclass[%
reprint,
superscriptaddress,
nofootinbib,
 amsmath,amssymb,
 aps,
showkeys,
]{revtex4-1}
\usepackage{graphicx}
\usepackage{subfigure}
\usepackage{dcolumn}
\usepackage{bm}

\usepackage[%
bookmarksnumbered,
bookmarksopen,
colorlinks,
citecolor=blue,
linkcolor=blue,
]{hyperref} 

\def\es0{$E_{sym}(\rho_0)$~}
\def\us0{$U_{sym}^{\infty}(\rho_{0})$}

\begin{document}

\title{Proton and kaon production in Au+Au collisions at $\sqrt{s_{\rm NN}}=3$~GeV}
\author{Shuang-Jie Liu}
\affiliation{School of Physics and Electronic Science, Guizhou Normal University, Guiyang
 550025, China}
\author{Gao-Feng Wei}
\email{Corresponding author: wei.gaofeng@gznu.edu.cn}
\affiliation{School of Physics and Electronic Science, Guizhou Normal University, Guiyang
 550025, China}
\author{Yu-Liang Zhao}
\affiliation{School of Physics and Electronic Science, Guizhou Normal University, Guiyang
 550025, China}
\author{Feng-Chu Zhou}
\affiliation{School of Physics and Electronic Science, Guizhou Normal University, Guiyang
550025, China}
\author{Zhen Wang}
\affiliation{School of Physics and Electronic Science, Guizhou Normal University, Guiyang
550025, China}

\begin{abstract}
Within an extended isospin- and momentum-dependent Boltzmann-Uehling-Uhlenbeck transport model, we study the protons, $K^+$ mesons and $\Lambda$ hyperons production in Au+Au collisions at $\sqrt{s_{\rm NN}}=3$~GeV. For the collision in 0-10\% centrality, we study the transverse momentum spectra and rapidity dependent mean transverse momentum for protons. For the collision in 10-40\% centrality, we study the directed and elliptic flows for protons and $K^+$ mesons. The results show that the momentum-dependent nuclear mean field with an incompressibility $K_0=230$~MeV can fit fairly the STAR experimental data, while the momentum-independent nuclear mean field with both $K_0=230$~MeV and $K_0=380$~MeV can only partially describe the experimental results. In addition, we also study the directed and elliptic flows for the associated $\Lambda$, observations reveal the same conclusions as for kaons. These findings indicate that the momentum dependence of nuclear mean field plays a significant role in understanding nuclear matter properties in heavy-ion collisions at $\sqrt{s_{\rm NN}}=3$~GeV.
\end{abstract}

\keywords{Momentum dependence of nuclear mean field, Collective flows, Kaon production}

\maketitle
\section{introduction}\label{introduction}
Heavy-ion collisions (HICs) provide a crucial way for studying strongly interacting matter under extreme conditions, enabling the production of high-temperature and high-density environments in the laboratory to simulate matter states such as those inside neutron stars or those at microseconds after the Big Bang. Studying the equation of state (EoS) of nuclear matter under such extreme conditions has long been one of the central topics in both nuclear physics and astrophysics~\cite{Latt16,Lim18} since it helps to understand the properties of dense nuclear matter and the quark-gluon plasma (QGP) matter~\cite{Latt05,Li08,Tews18,Munz16,MA20}. 

In HICs, the azimuthal distribution of final state particles in momentum space is usually analyzed through the Fourier expansion, in which the first and second coefficients of the expansion terms correspond to the directed flow $v_1 = \langle \cos \phi \rangle$ and elliptic flow $v_2 = \langle \cos 2\phi \rangle$, where $\phi$ is the azimuthal angle of an outgoing particle relative to the reaction plane~\cite{Posk98,Nara22}. These flow observables directly reflect the pressure and/or pressure gradient generated in HICs~\cite{Dan98,Russ06,Zhou25}, exhibiting high sensitivity to the nuclear EoS, and are widely used to extract key nuclear EoS information through comparing to the experimental data~\cite{Heinz13,Dan02,Stock86,Pan93,Zhang94,Aam10,Adam14,Andr05,Chan97,Cser99,Ali23,Feng24}. However, the nuclear EoS at high densities is still significantly uncertain. For example, as indication in Refs.~\cite{Gale87,Isse05,Zhou25,Nara99,Weil16}, it seems that both the momentum-dependent soft EoS and momentum-independent stiff EoS can all reproduce the experimental data. Therefore, this naturally leads us to study the incompressibility $K_0$$-$the coefficient of the leading term in the expansion of nuclear EoS$-$and thus to advance an understanding of the nuclear EoS.

Recently, the high precision data at $\sqrt{s_{\rm NN}}=3$~GeV published by the STAR Collaboration provide an important opportunity to investigate the nuclear EoS at 3-4 times the saturation density~\cite{M.S22,M.I24}. However, it is worth repeating again that the nuclear EoS extracted from this experiment still has considerable uncertainty. For example, as indication in Refs.~\cite{M.S22,M.K24}, it seems that the momentum-independent nuclear mean field with both soft and stiff EoS could fit this experimental data, while in Ref.~\cite{Zhou25} only the momentum-dependent mean field with a soft EoS could provide a good fit to this experimental data. To this situtation, it is naturally need to further examine effects of the momentum dependence of nuclear mean field in HICs in this experiment. To this end, based on an extended isospin- and momentum-dependent Boltzmann-Uehling-Uhlenbeck transport (IBUU) model, we study the proton and kaon production as well as the associated $\Lambda$ production in Au + Au collisions at $\sqrt{s_{\rm NN}}=3$~GeV. The focus is on whether the momentum-dependent nuclear mean field with an incompressibility $K_0=230$ MeV can describe the STAR data at $\sqrt{s_{\rm NN}}=3$~GeV. For comparison, we also adopt the momentum-independent nuclear mean field with a soft EoS ($K_0$=230 MeV) and a stiff EoS ($K_0$=380 MeV) in our model. The results show as in Ref.~\cite{Zhou25} that the momentum-dependent nuclear mean field  with the incompressibility $K_0=230$ MeV indeed could fairly describe the STAR experimental data at $\sqrt{s_{\rm NN}}=3$~GeV, while the momentum-independent nuclear mean field can only partially describe the STAR data. These findings indicate that the momentum dependence of nuclear mean field plays a significant role in understanding nuclear matter properties in HICs at $\sqrt{s_{\rm NN}}=3$~GeV.

\section{The Model}\label{Model}
This study is carried out within an IBUU transport model~\cite{Das03,li041,Chen05}. However, to simulate the HICs at higher energy, e.g, at $\sqrt{s_{\rm NN}}=3$~GeV, we have upgraded the model by including more reation channels and the corresponding particles, e.g, $N^{*}$(1535), $\eta$, $K$, $\Lambda$, $\Sigma$ etc., in which the production of $K$, $\Lambda$, and $\Sigma$ as well as the corresponding reaction channels have been included in our recent study~\cite{wei24}. As for the newly included  $N^{*}$(1535) and $\eta$ as well as the corresponding reaction channels, we will briefly introduce them later in the following. 
As a default case, the momentum-dependent nuclear mean field (labelled as MDI) used in the model is expressed as,
\begin{eqnarray}
	U(\rho,\delta ,\vec{p},\tau ) &=&A_{u}\frac{\rho _{-\tau }}{\rho _{0}}%
	+A_{l}\frac{\rho _{\tau }}{\rho _{0}}\notag \\
	&+&B{\Big(}\frac{\rho}{\rho _{0}}{\Big)}^{\sigma }(1-x\delta^2)-4\tau x\frac{B}{\sigma+1}\frac{\rho^{\sigma-1}}{\rho_{0}^\sigma}\delta\rho_{-\tau}
	\notag \\
	&+&\frac{2C_{l }}{\rho _{0}}\int d^{3}p^{\prime }\frac{f_{\tau }(%
		\vec{p}^{\prime })}{1+(\vec{p}-\vec{p}^{\prime })^{2}/\Lambda ^{2}}
	\notag \\
	&+&\frac{2C_{u }}{\rho _{0}}\int d^{3}p^{\prime }\frac{f_{-\tau }(%
		\vec{p}^{\prime })}{1+(\vec{p}-\vec{p}^{\prime })^{2}/\Lambda ^{2}},
	\label{IMDIU}
	\label{IMDIU}
\end{eqnarray}%
where $\tau=1$ for neutrons and $-1$ for protons, and $A_{u}$, $A_{l}$, $C_{u}(\equiv C_{\tau,-\tau})$ and $C_{l}(\equiv C_{\tau,\tau})$ are expressed as
\begin{eqnarray*}
	A_{l}&=&A_{l0}+U_{sym}^{\infty}(\rho_{0}) + x\frac{2B}{\sigma+1}\notag,\\
	A_{u}&=&A_{u0}-U_{sym}^{\infty}(\rho_{0}) - x\frac{2B}{\sigma+1}\notag,\\
	C_{l}&=&C_{l0}-2\big{(}U_{sym}^{\infty}(\rho_{0})-2z\big{)}\frac{p_{f0}^{2}}{\Lambda^{2}\ln \big{[}(4p_{f0}^{2}+\Lambda^{2})/\Lambda^{2}\big{]}},\\
	C_{u}&=&C_{u0}+2\big{(}U_{sym}^{\infty}(\rho_{0})-2z\big{)}\frac{p_{f0}^{2}}{\Lambda^{2}\ln \big{[}(4p_{f0}^{2}+\Lambda^{2})/\Lambda^{2}\big{]}}.
\end{eqnarray*}
The values of eight parameters embedded in the above expressions, i.e., $A_{l0}$, $A_{u0}$, $B$, $\sigma$, $C_{l0}$, $C_{u0}$, $\Lambda$, and $z$, are connected to the properties of nuclear matter at $\rho_{0}=0.16$~fm$^{-3}$. Among them, the first seven parameters are: $A_{l0}=A_{u0}=-66.963$~MeV, $B=141.963$~MeV, $C_{l0}=-60.486$~MeV, $C_{u0}=-99.702$~MeV, $\sigma=1.2652$, and $\Lambda=2.424p_{f0}$, where $p_{f0}$ is the nucleon Fermi momentum in symmetric nuclear matter (SNM) at $\rho_{0}$. The eighth parameter $z$ is associated with the symmetry energy parameter $x$ that is used to mimic the slope value $L\equiv{3\rho({dE_{sym}}/d\rho})$ of symmetry energy at $\rho_{0}$. Because we aim in this study to extract the nuclear EoS, we therefore use a certain value of $L=62.47$~MeV and the corresponding parameters $x$ and $z$ are $x=0$ and $z=0$. In this case, the properties of nuclear matter at $\rho_{0}=0.16$~fm$^{-3}$ are fixed including the binding energy $-16$~MeV, the pressure $P_{0}=0$~MeV/fm$^{3}$, the incompressibility $K_{0}=230$~MeV for SNM at $\rho_{0}$, the isoscalar effective mass $m^{*}_{s}=0.7m$, the isoscalar potential at infinitely large nucleon momentum $U^{\infty}_{0}(\rho_{0})=75$~MeV, the isovector potential at infinitely large nucleon momentum $U^{\infty}_{sym}(\rho_{0})=-100$~MeV as well as the symmetry energy $E_{sym}(\rho_{0})=32.5$~MeV and $E_{sym}(2\rho_{0}/3)=24.75$~MeV.  

For comparison, we also employ a  commonly used momentum-independent mean field~\cite{Bertsch88,Baran05,Li08} (labelled as MID) in form of,
\begin{eqnarray}	
	U(\rho,\delta,\tau)=\alpha\left(\frac{\rho}{\rho_0}\right)+\beta\left(\frac{\rho}{\rho_0}\right)^\xi+\nu_{asy}(\rho,\delta,\tau),
\end{eqnarray} 
in which, the third term, i.e., $\nu_{asy}(\rho, \delta, \tau )$, expressed as~\cite{Li02},
\begin{equation}\begin{aligned}
		\nu_{asy}(\rho,\delta,\tau) & =\tau\left(E_{sym}(\rho_{0})u^{\gamma}-12.7u^{2/3}\right)\delta \\
		& +\left(E_{sym}(\rho_{0})(\gamma-1)u^{\gamma}+4.2u^{2/3}\right)\delta^{2},
\end{aligned}\end{equation}
is exactly the isovector part we considered in the momentum-independent mean filed that results in the symmetry energy with a form $E_{sym}(\rho)=E_{sym}(\rho_0)u^{\gamma}$ ,
where $u\equiv\rho/\rho_{0}$ is the reduced density. The parameter $\gamma$ controls the stiffness of the symmetry energy. To ensure consistency, we also adopt $L=62.47$ MeV (corresponding to $\gamma$=0.64) as the symmetry energy slope in the MID scenario. The stiffness of the nuclear EoS is governed by parameters $\alpha$, $\beta$, and $\xi$. For comparison, we consider a stiff EoS ($K_0=380$~MeV) with parameters $\alpha=-123.03$ MeV, $\beta=69.77$~MeV, $\xi=2.01$ and a soft EoS ($K_0=230$~MeV) with parameters $\alpha=-232.15$~MeV, $\beta=178.89$~MeV, $\xi=1.30$.

\begin{figure}[htbp!]
	\includegraphics[width=\columnwidth]{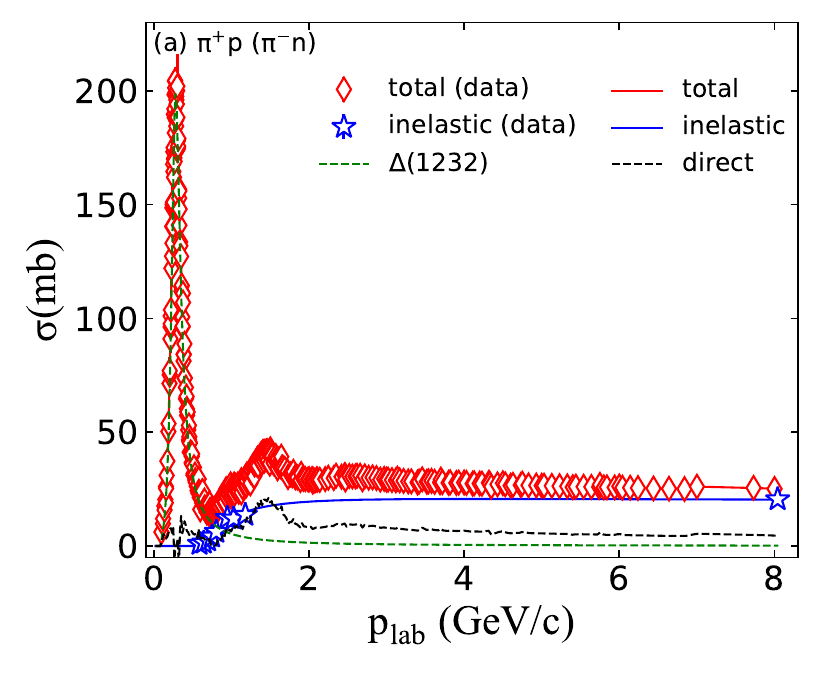}
	\includegraphics[width=\columnwidth]{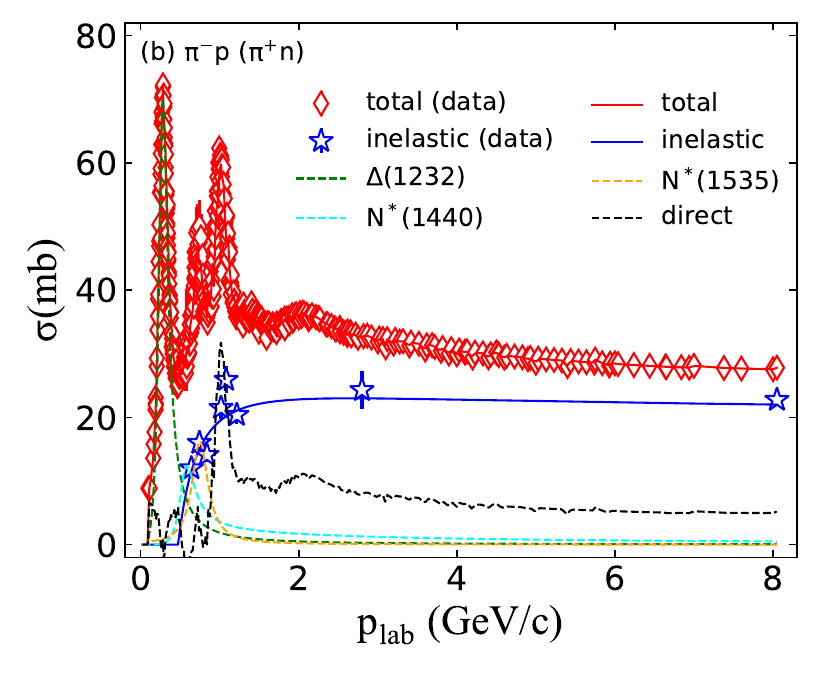}
	\includegraphics[width=\columnwidth]{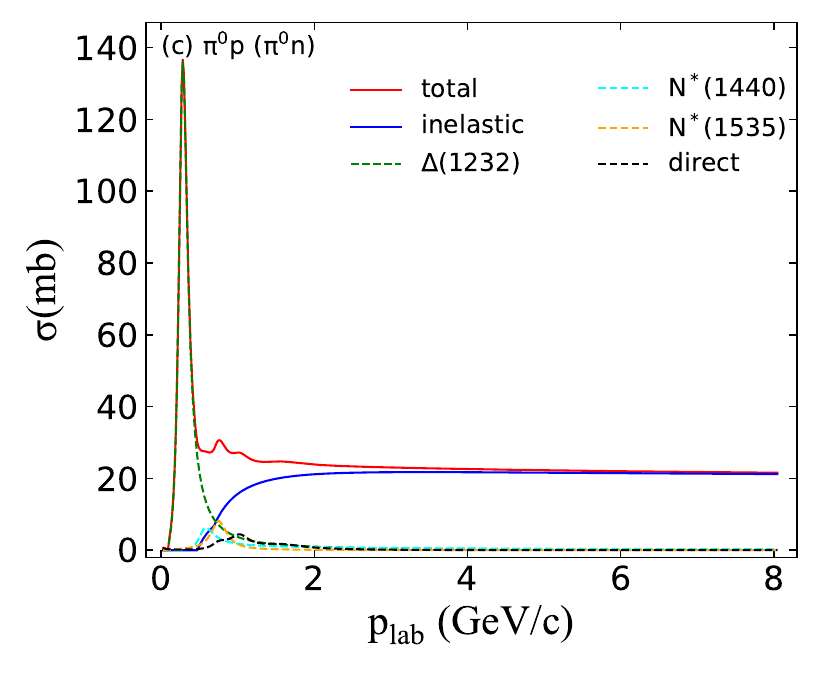}
	\caption{(Color online) The total, inelastic cross sections and the contributions from three lowest resonant states as well as that of a direct process as a function of $p_{lab}$. The symbols represent the available experimental data, and the lines represent the ones used in the model.}
	\label{np}
\end{figure}

On the other hand, for the HICs at higher collision energy, we notice that apart from the pion production channels, i.e., $NN\rightarrow{NN}+\pi$, the $\eta$ production channels are also important, i.e., $NN\rightarrow{NN}+\eta$. Therefore, except for the $\Delta$(1232) and $N^{*}$(1440) resonances in original IBUU model, we also include the production of $N^{*}$(1535) resonances and $\eta$ mesons in two types of most important reaction channels, i.e., nucleon-nucleon collisions and nucleon-pion as well as nucleon-$\eta$ collisions. As for the corresponding cross sections, we generally fit the experimental and/or empirical data. For example, for the production of a $\Delta$ and $N^{*}(1440)$ in a nucleon-nucleon collision, we determine the corresponding cross sections using the  parameterization formula introduced by VerWest and Arndt~\cite{VerWest80} as used in original BUU and ART models~\cite{li041,Chen05,Li95}, i.e.
\begin{eqnarray}
	\sigma(p+p&\rightarrow& n+\Delta^{++})=\sigma_{10}+\frac{1}{2}\sigma_{11},\\
	\sigma(p+p&\rightarrow&p+\Delta^+)=\frac{3}{2}\sigma_{11}\\
	\sigma(n+p&\rightarrow&p+\Delta^0)=\frac{1}{2}\sigma_{11}+\frac{1}{4}\sigma_{10},\\
	\sigma(n+p&\rightarrow&p+\Delta^+)=\frac{1}{2}\sigma_{11}+\frac{1}{4}\sigma_{10},\\
	\sigma(p+p&\rightarrow&p+N^{*+}(1440))=0,\\
	\sigma(n+p&\rightarrow&p+N^{*0}(1440))=\frac{3}{4}\sigma_{01},\\
	\sigma(n+p&\rightarrow&p+N^{*+}(1440))=\frac{3}{4}\sigma_{01},
\end{eqnarray}
where the parameterization formula $\sigma_{if}$ is defined as,
\begin{eqnarray}
	\sigma_{if}(\sqrt{s})=\frac{\pi(\hbar c)^{2}}{2p^{2}}\alpha\left(\frac{p_{r}}{p_{0}}\right)^{\beta}\frac{m_{0}^{2}\Gamma^{2}(q/q_{0})^{3}}{(s^{*}-m_{0}^{2})^{2}+m_{0}^{2}\Gamma^{2}},
\end{eqnarray}
and the parameters $\alpha, \beta, m_{0}, \Gamma$ and other quantities can be found in Ref.~\cite{VerWest80}. As to the $N^*(1535)$, considering that it has the approximate 50\% probability decaying into $\eta$ meson, we therefore determine the cross section for the production of $N^*(1535)$
resonance from the empirical $\eta$ production cross section~\cite{Wolf93} as used in ART model~\cite{Li95}, i.e.,
\begin{equation}
	\sigma(NN\to NN^*(1535))\approx2\sigma(NN\to NN\eta).
\end{equation}
Specifically, taking the production of a $N^*(1535)$ in a proton-proton collision as an example, the corresponding cross section could be determined as
\begin{eqnarray}
	\sigma(pp\to pN^{*+}(1535))&\approx& 2\sigma(pp\to pp\eta) \notag \\
    &\approx& 2\frac{0.102(\sqrt{s}-2.424)}{0.058+(\sqrt{s}-2.424)^{2}}.
\end{eqnarray}

For the pion-nucleon and/or $\eta$-nucleon collisions, the corresponding cross sections of producing the $\Delta(1232)$, $N^{*}(1440)$ and $N^{*}(1535)$ are evaluated by the Breit-Wigner formula as used in original BUU and/or ART models~\cite{li041,Chen05,Li95}. For example, for the pion-nucleon collisions, we determine the cross sections as,
\begin{equation}
	\sigma_{\pi N\to R}(\sqrt{s})=\sigma_{max}(|\mathbf{p}_0/\mathbf{p}|)^2\frac{0.25\Gamma^2(\mathbf{p})}{0.25\Gamma^2(\mathbf{p})+(\sqrt{s}-m_0)^2},
\end{equation}
where $m_0$ denotes the resonance ($R$) mass centroid, and the $p$ and $p_0$ represent the pion momenta at energies $\sqrt{s}$ and $m_0$, respectively. The maximum cross section $\sigma_{max}$ is determined by fitting to the available experimental total cross section data, i.e., 190mb, 130mb, and 67mb for $\pi^{+}+p\to\Delta^{++}(\pi^{-}+n\to\Delta^{-})$, $\pi^{0}+p\to\Delta^{+}(\pi^{0}+n\to{\Delta}^{0})$ and $\pi^{-}+p\to\Delta^{0}(\pi^{+}+n\to\Delta^{+})$, respectively. And 12mb, 6mb, 16mb, 8mb for $\pi^{-}+p\to N^{*0}(1440)(\pi^{+}+n\to N^{*+}(1440))$, $\pi^{0}+p\to N^{*+}(1440)(\pi^{0}+n\to N^{*0}(1440))$, $\pi^{-}+p\to N^{*0}(1535)(\pi^{+}+n\to N^{*+}(1535))$, $\pi^{0}+p\to N^{*+}(1535)(\pi^{0}+n\to N^{*0}(1535))$, respectively. Also, we use 74mb for $\eta+p\to N^{*+}(1535)(\eta+n\to N^{*0}(1535)$ as used in ART model~\cite{Li95}. Fig.~\ref{np} presents a comparison of the used total and inelastic cross sections for pion-nucleon collisions in the model with the available experimental data~\cite{Olive14,PDG88}. Apart from the inelastic cross sections and the contributions from three lowest resonant states, the total cross sections used in the model also incorporate the contributions from a direct process denoted as "direct" as shown in Fig.~\ref{np}. The cross section of this direct process, i.e., $\pi+N\to{\pi+N}$ that is used to simulate the higher but shorter-lived resonances, is calculated by subtracting the sum of inelastic cross sections and the contributions of three lowest resonant states from the experimental total cross sections.
\begin{figure*}[hbt]
	\includegraphics[width=0.92\textwidth,height=0.34\textheight]{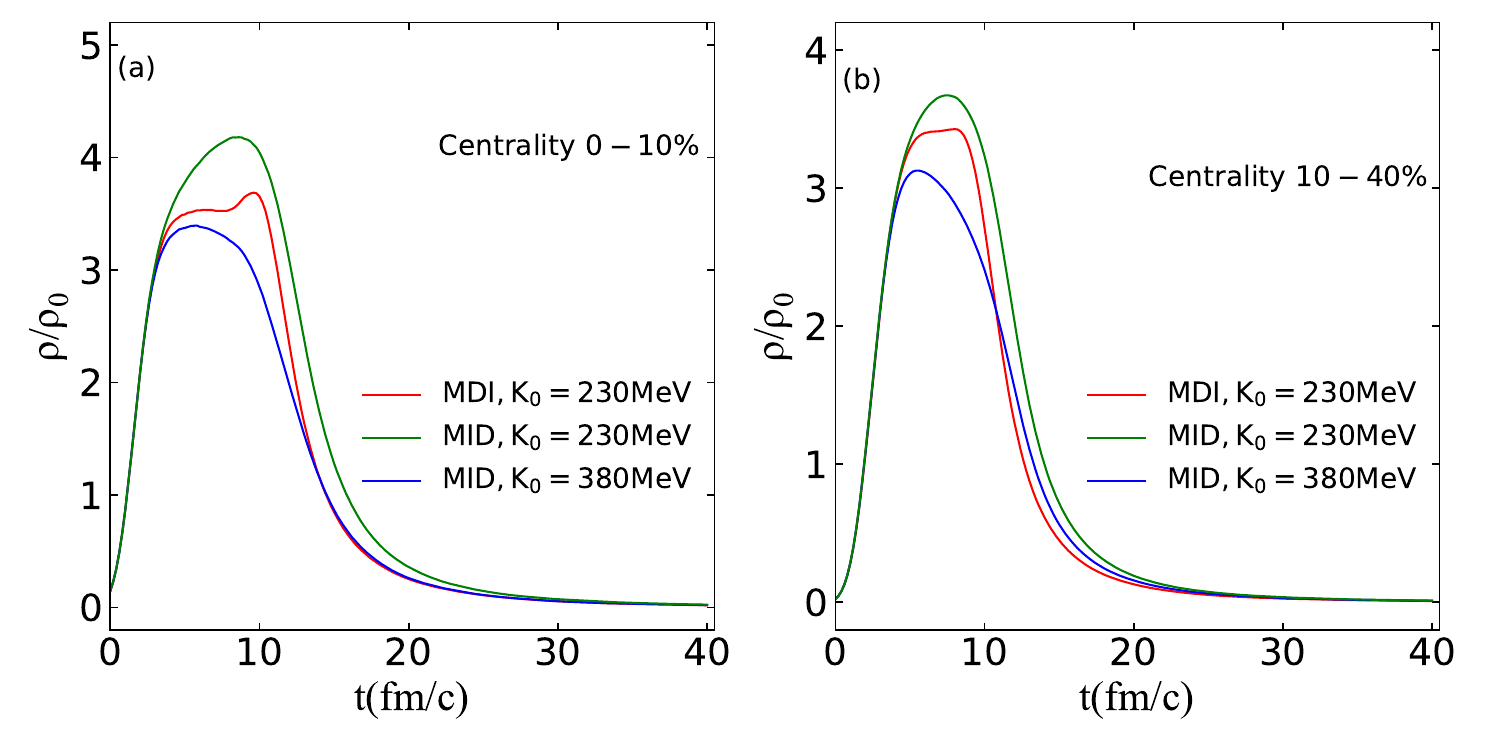}
	\caption{(Color online) Evolution of central region reduced
density $\rho$/$\rho_0$ in 0-10\% (a) and 10-40\% (b) Au+Au collisions
at  $\sqrt{s_{\rm NN}}=3$~GeV.} 
	\label{den}
\end{figure*}
\begin{figure*}[ht]
	\includegraphics[width=0.92\textwidth,height=0.34\textheight]{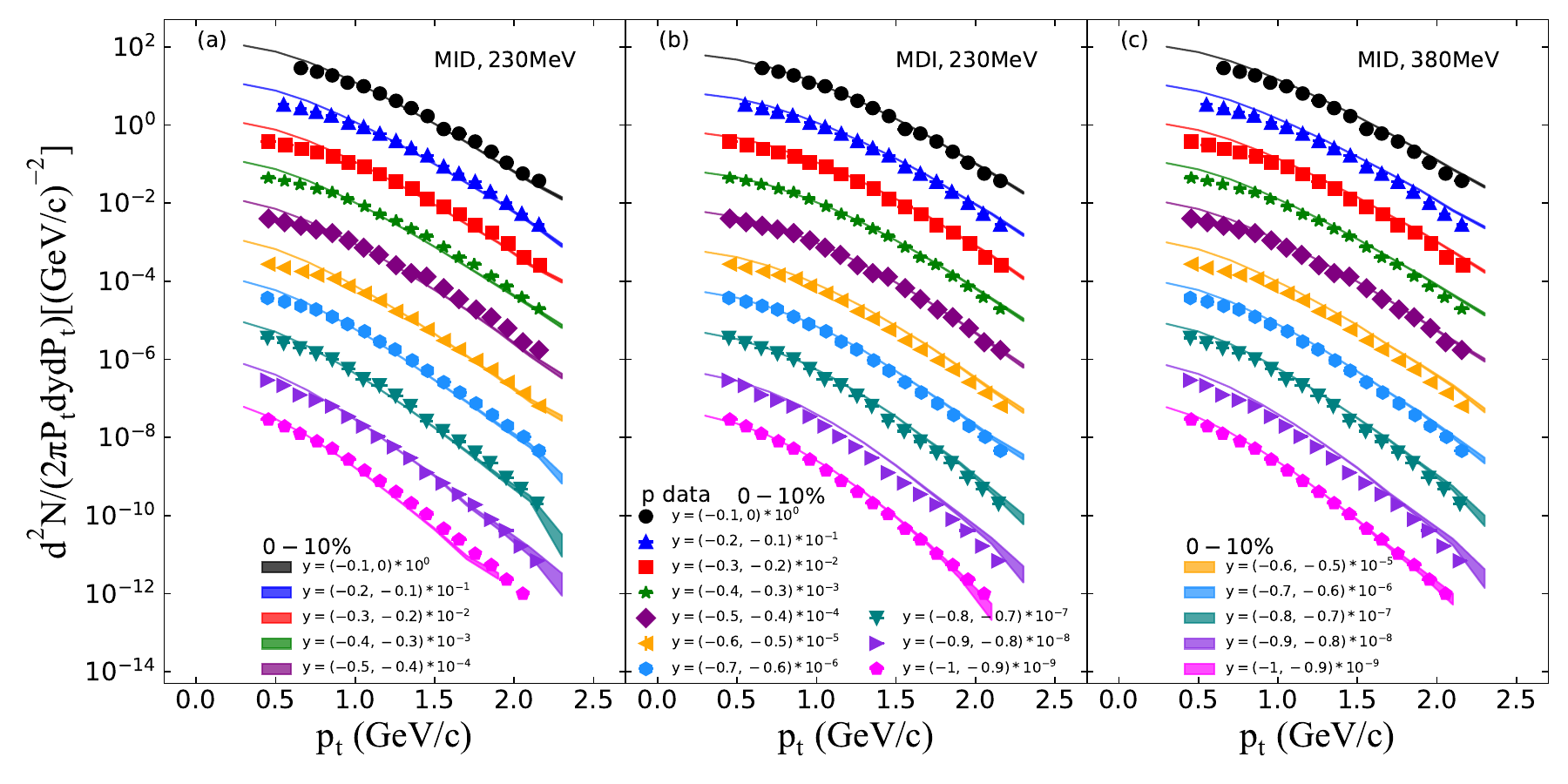}
	\caption{(Color online) Proton transverse momentum ($p_t$) spectra from theoretical simulations at different rapidity intervals in 0-10\% Au+Au collisions at $\sqrt{s_{\rm NN}}=3$~GeV in comparison with the corresponding STAR data~\cite{M.I24}.} 
	\label{pt}
\end{figure*}
\begin{figure}[htbp!]
	\includegraphics[width=\columnwidth]{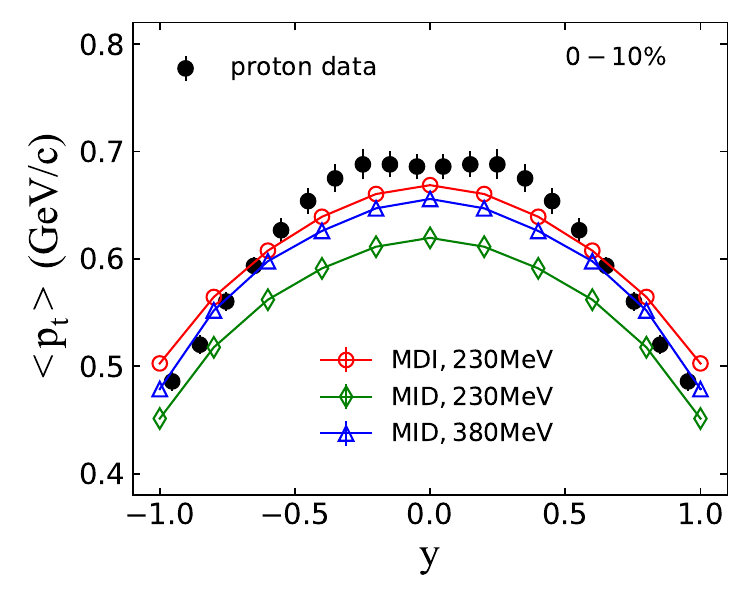}
	\caption{(Color online)Proton mean transverse momentum ($\langle p_t \rangle$) as a function of the rapidity in 0-10\% Au+Au collisions at  $\sqrt{s_{\rm NN}}=3$~GeV in comparison with the corresponding STAR data~\cite{M.I24}. }
	\label{pty}
\end{figure}

\section{Results and Discussions}\label{Results and Discussions}
Now, we present the results for Au+Au collisions at $\sqrt{s_{\rm NN}}=3$~GeV. To exclude uncertainties of statistical nature from our physical considerations, we have simulated $1 \times 10^{5}$ events for the collisions with 0-10\% centrality, while for the collisions with 10-40\% centrality, we have performed $6 \times 10^{5}$ events to meet the higher statistical requirements of the collective flow analysis.
 
To better understand the effects of the incompressibility $K_0$ and the momentum dependence of nuclear mean field on the reaction observables, we first show the attainable density in the central region during the reaction and thus to have a global picture on the reaction dynamics. Fig.~\ref{den}(a) and Fig.~\ref{den}(b) show the time evolution of the central compression density for Au+Au collisions in 0-10\% and 10-40\% centrality, respectively. Obviously, due to a smaller $K_{0}$ value usually corresponds to a softer EoS that could cause an easier compression in the central region of heavy-ion collisions, we therefore could find the attainable density with the incompressibility $K_0=230$ MeV is significantly larger than that with the $K_0=380$ MeV within the MID nuclear mean field. However, it is interesting to see that the compression density with the MDI nuclear mean field with the incompressibility $K_0=230$~MeV falls between these two cases of MID scenarios. This observation implies that the nuclear compression is significantly affected by both the bulk EoS and the momentum dependence of nuclear mean field. In addition, It is worth mentioned that the increase of $K_0$ value and the consideration of momentum dependence of mean field exhibit similar effects in the nuclear compression. Naturally, this characteristic will inevitably be reflected by the collective flows of nucleons.

\begin{figure}[htbp!]
	\includegraphics[width=\columnwidth]{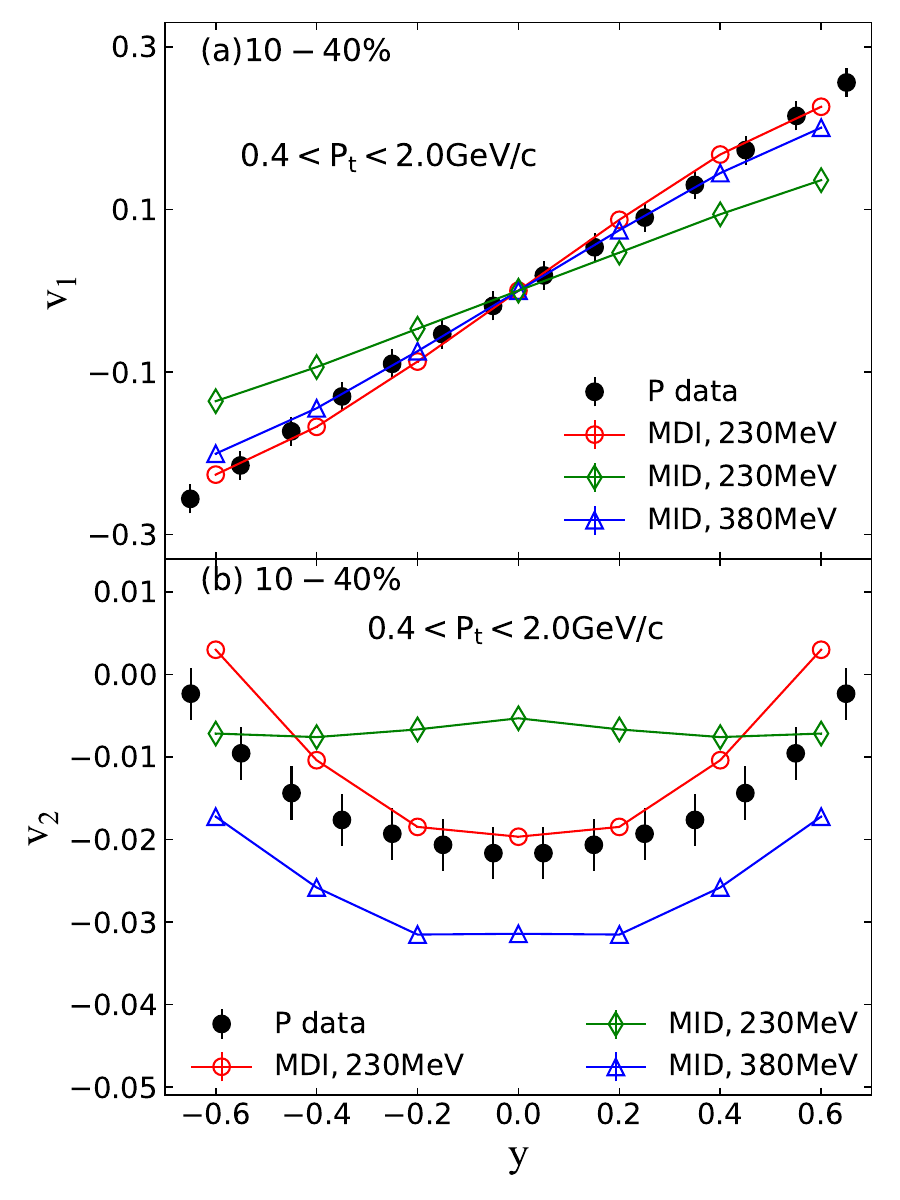}
	\caption{(Color online) The directed (a) and elliptic (b) flows of protons as a function of the rapidity in 10-40\% Au+Au collisions at $\sqrt{s_{\rm NN}}=3$~GeV in comparison with the corresponding STAR data~\cite{M.S22}.}
	\label{v1v2y}
\end{figure}

\subsection{Proton flow}
Fig.~\ref{pt} shows the simulation results of proton transverse momentum spectra for different rapidity intervals in 0-10\% Au+Au collisions at $\sqrt{s_{\rm NN}}=3$~GeV, in comparison with the corresponding STAR experimental data \cite{M.I24}. It can be observed that all three cases could fairly reproduce the STAR experimental data. To this situation, one naturally wonders whether these three cases could equally reproduce other observables. To answer this question, we present in Fig.~\ref{pty} the mean transverse momentum as a function of the rapidity in the same reaction. First, it is clear to see that the simulation results with both the MDI mean field with the $K_0=230$ MeV and the MID mean field with the $K_0=380$ MeV could reasonably fit the experimental data, while those with the MID mean field with the $K_0=230$ MeV underestimate the proton transverse momentum. Second, compared with the MID mean field with the incompressibility $K_0=230$ MeV, the MID mean field with the $K_0=380$ MeV causes a larger amplitude for the proton mean tranverse momentum, this is due to the stiff EoS generating stronger pressure and/or pressure gradient and thus enhancing the transverse emission of protons as indicated in Ref.~\cite{Russ06}. Moreover, we can also find that the proton mean transverse momentum with the MDI mean field with the $K_0=230$ MeV is significantly larger than that with the MID mean field with the $K_0=230$ MeV, demonstrating the enhancement effects from the momentum dependence of the nuclear mean field, this is also consistent with the indication in Refs.~\cite{Zhou25,Kire24}.

To further explore the impact of the mean-field momentum dependence on collective flows, we show in Fig.~\ref{v1v2y} the rapidity dependent directed and elliptic flows for protons in 10-40\% Au+Au collisions at $\sqrt{s_{\rm NN}}=3$~GeV in comparison with the corresponding STAR experimental data \cite{M.S22}. First, consistent with the indication in many previous studies~\cite{Pan93,Zhang94,Hill20,Chen03,fang23} that the momentum-independent stiff EoS could reproduce the transverse/directed flow data from HICs equally well as the momentum-dependent soft EoS, we can find that both the MDI mean field with the $K_0=230$ MeV and the MID mean field with the $K_0=380$ MeV could fairly fit the proton directed flow data, while the MID mean field with the $K_0=230$ MeV underestimates the amplitude of proton directed flow as shown in Fig.~\ref{v1v2y}(a). However, as shown in Fig.~\ref{v1v2y}(b), the MID mean field with the stiff EoS ($K_0=380$ MeV) overestimates the amplitude of proton elliptic flows, while the soft EoS ($K_0=230$ MeV) underestimates the amplitude of proton elliptic flows. In contrast, only the MDI mean field with the $K_0=230$ MeV could fairly reproduces the proton elliptic flow data. This indicates that although the increase of $K_0$ value or the consideration of momentum dependence of mean field, respectively, could cause approximately equal effects on the proton directed flows, their influence on the elliptic flow is different. As pointed out in the literatures~\cite{Chen18,Nara99,Zhang18}, softening of the equation of state suppresses the squeeze-out effect, thus leading to an enhancement of elliptic flow. This trend is also observed in the present study, e.g., for the MID mean field, increasing $K_0$ from 230 MeV to 380 MeV stiffens the equation of state, resulting in a larger pressure and/or pressure gradient and thus a more negative elliptic flow. In contrast, the pressure and/or pressure gradient generated by the MDI mean field with the $K_0=230$ MeV is enhanced but still remains at a moderate level, thus providing a fair description of the elliptic flow data. Therefore, the momentum dependence of mean field remains crucial for interpreting the experimental data in HICs at $\sqrt{s_{\rm NN}}=3$~GeV especially for the proton elliptic flow data.

\subsection{$K^+$ production}
It is well known that kaon observable is a more promising probe for the study of nuclear EoS in HICs~\cite{Aiche85,Fuch01,Hart12}. Actually, considering that the $K^{+}$ and the associated $\Lambda$ are mainly produced from nucleon-nucleon collisions (e.g., $NN\to N \Lambda K^{+}$) and their collisions with nucleons, naturally, the influence of momentum dependence on proton collective flow may also be reflected in the collective flow of $K^{+}$ and $\Lambda$.  
Therefore, we turn to in this section examine the $K^{+}$ and the $\Lambda$ production in 10-40\% Au+Au collisions at $\sqrt{s_{\rm NN}}=3$~GeV that has been carried out by the STAR Collaboration~\cite{M.S22}. Recently, we have already investigated the production mechanism of $K^+$ mesons in Au + Au collisions at $\sqrt{s_{\rm NN}}=2.4$~GeV~\cite{wei24}, in which we found that the medium modification of kaon mass plays a significant role in kaon production in HICs, and the two commonly used kaon potential scenarios, i.e., empirical scattering length and chiral Lagrangian scenarios, are almost equally good in kaon production in HICs. Therefore, in this study we employ the empirical kaon-nucleon scattering length scenario, i.e.,
\begin{equation}
    \omega_{K}={\Big[}m^{2}_{K}+{\mathbf k}^{2}-4\pi\alpha_{KN}{\big(}1+\frac{m_{K}}{{m_{N}}}{\big)}\rho {\Big]}^{1/2},
\end{equation}
where $m_{K}$ and $m_{N}$ are the kaon and nucleon masses in free space, $\bf{k}$ is the kaon momentum, $\rho$ is the baryon density, and $\alpha_{KN}{\approx}-0.255$ fm is the kaon-nucleon scattering length that leads to a repulsive kaon potential of 30 MeV at $\rho_{0}$.
Correspondingly, the kaon potential could be determined as $U_{K}=\omega_{K}-(m^{2}_{K}+{\bf k}^{2})^{1/2}$. Apart from the kaon potential, the kaon effective mass in dense nuclear medium is also a key factor for kaon production in HICs, we therefore also consider the medium modification of kaon mass, i.e., $m_{K}^{*}={\big[}m^{2}_{K}-4\pi\alpha_{KN}{\big(}1+\frac{m_{K}}{{m_{N}}}{\big)}\rho {\big]}^{1/2}$, in simulations of kaon production in Au + Au collisions at $\sqrt{s_{\rm NN}}=3$~GeV. For the details of kaon production in HICs, we refer readers to see Ref.~\cite{wei24}.
\begin{figure}[htbp!]
	\includegraphics[width=\columnwidth]{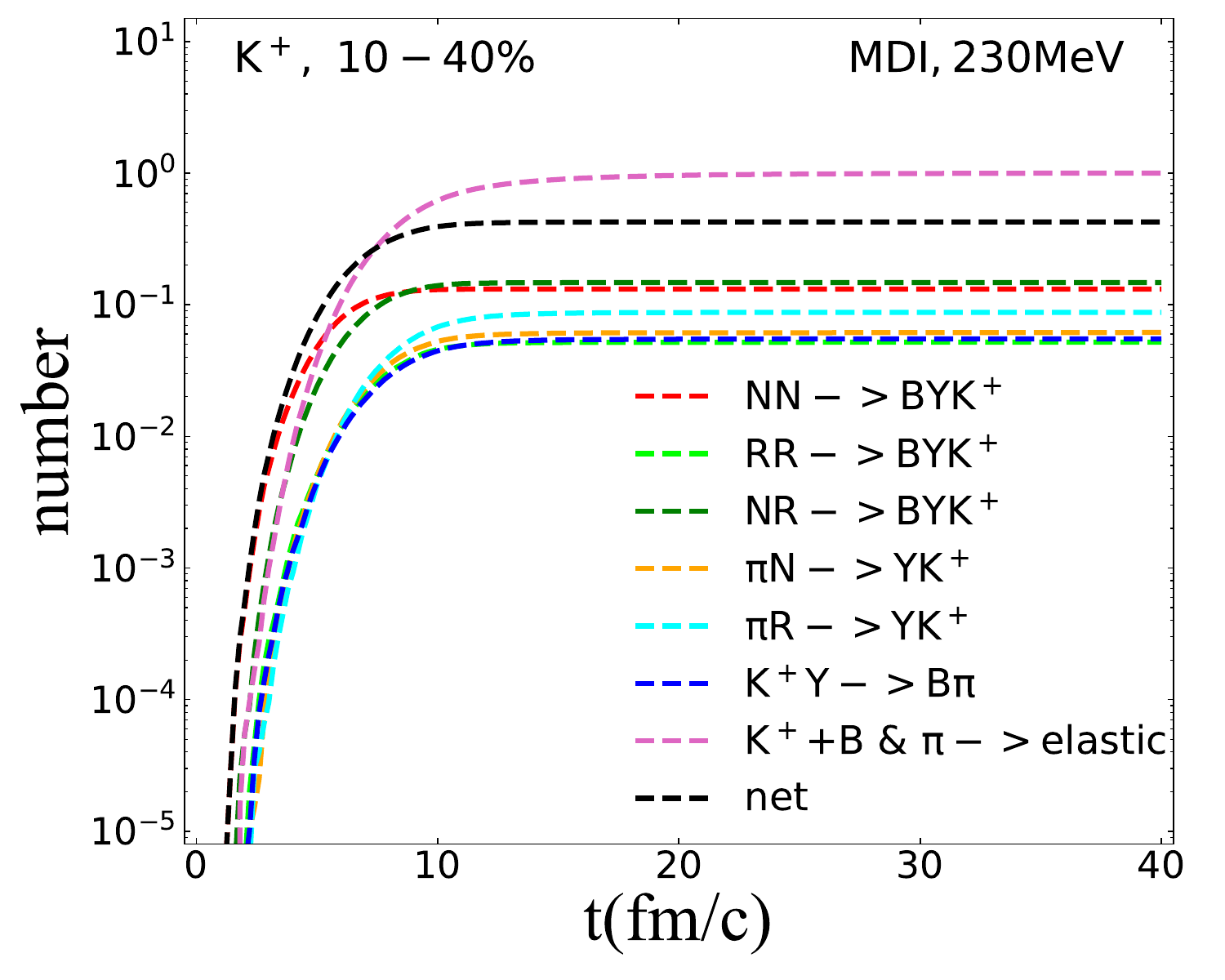}
	\caption{(Color online)Evolution of the kaon number from various channels in 10-40\% Au+Au collisions at $\sqrt{s_{\rm NN}}=3$~GeV. The symbol $N$ denotes nucleons, $R$ denotes the resonances, $Y$ is the $\Lambda$ or $\Sigma$, and $B$ is the baryon, i.e., nucleons or resonances.}
	\label{kaon}
\end{figure}

Shown in Fig.~\ref{kaon} is the evolution of kaon number from different channels in the reaction with the MDI nuclear mean field with the incompressibility $K_0=230$ MeV. First, we can see that elastic collisions between kaons and baryons as well as pions are dominant compared to all of inelastic collisions. Second, it can be seen that the nucleon-resonance (NR) and nucleon-nucleon (NN) channels emerge as the top two contributors to $K^+$ yields. These features are consistent with those observed in simulations of the HADES Au + Au collisions at $\sqrt{s_{\rm NN}}=2.4$~GeV~\cite{wei24}, demonstrating that the production mechanism of kaons in HICs at $\sqrt{s_{\rm NN}}=3$~GeV is the same as those at low energies in HICs.

\begin{figure}[htbp!]
	\includegraphics[width=\columnwidth]{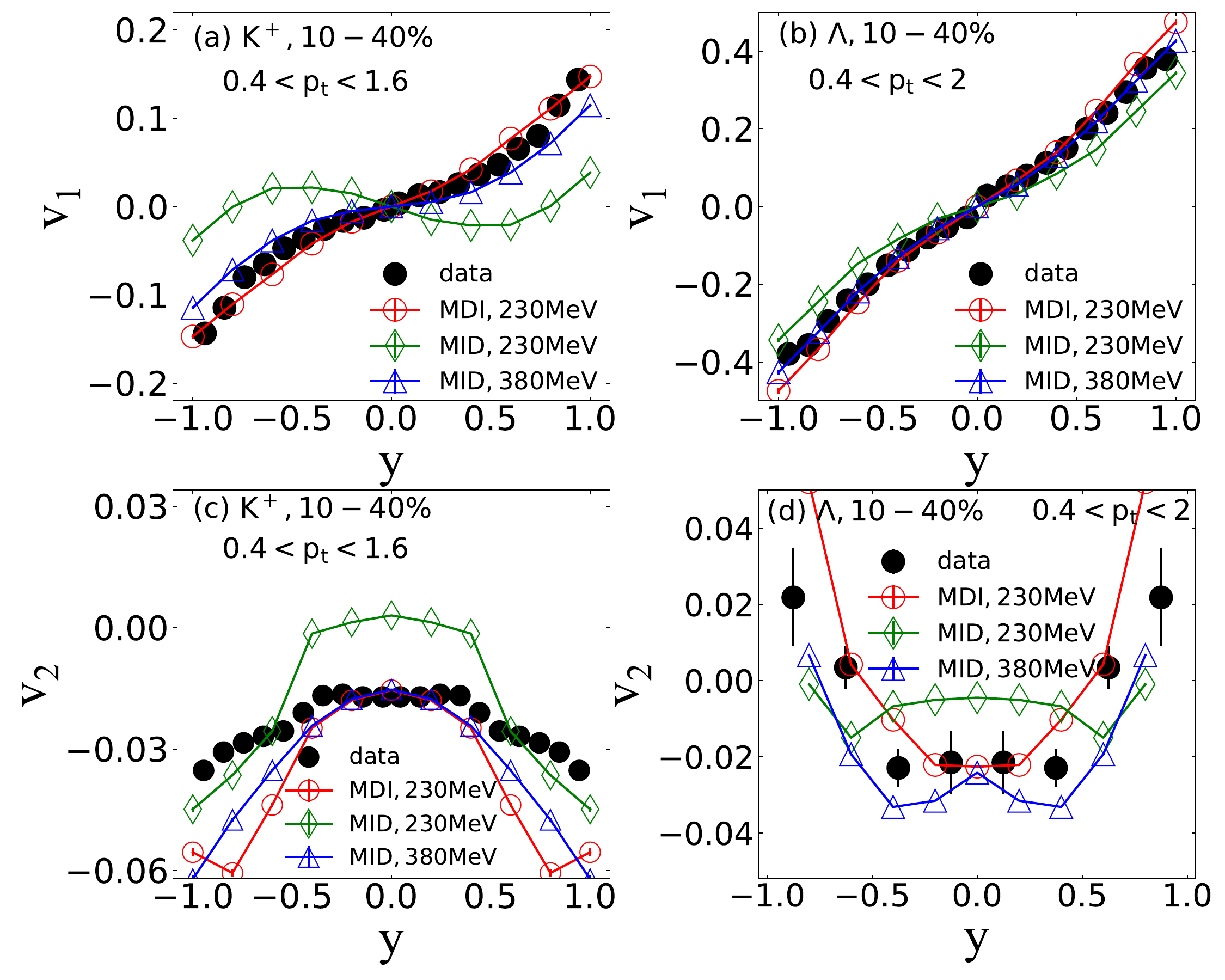}
	\caption{(Color online)The directed and elliptic flows of $K^{+}$ and  $\Lambda$ in 10-40\% Au + Au collisions
at $\sqrt{s_{\rm NN}}=3$~GeV as a function of the rapidity in comparison with the STAR data~\cite{M.S22}.}
	\label{kaonslamda}
\end{figure}

Shown in Fig.~\ref{kaonslamda}(a) and Fig.~\ref{kaonslamda}(c) are the rapidity dependent directed and elliptic flows for $K^+$ mesons. It is seen that the MID nuclear mean field with the incompressibility $K_0=380$ MeV causes a larger amplitude for the kaon flow compared with the MID nuclear mean field with the incompressibility $K_0=230$ MeV, reflecting a stronger repulsive interaction between the kaons and nucleons. Again, it seems that both the MDI nuclear mean field with the incompressibility $K_0=230$ MeV and the MID nuclear mean field with the incompressibility $K_0=380$ MeV can fairly fit the kaon flow data, certainly, the discrepancies are still visible at large rapidities for the elliptic flows. This naturally leads us to look at the associated $\Lambda$ flow to further check this observation as shown in Fig.~\ref{kaonslamda}(b) and Fig.~\ref{kaonslamda}(d), where the $\Lambda$ potential is calculated according to the  quark counting rule, i.e., $U_{\Lambda} = 2/3(1/3U_n + 2/3U_p)$~\cite{Mosz74,Chung01,Yong22}. Indeed, it is seen that the collective flows of $\Lambda$ share similar feature as that of kaons. However, we can clearly find that only the MDI nuclear mean field with the incompressibility $K_0=230$ MeV can simultaneously present a fair fit to the $\Lambda$ directed and elliptic flow data. This indicates again that the momentum dependence of nuclear mean field plays a vital role in understanding the nuclear matter properties and describing experimental data of HICs even at high energies.

Before ending this part, we give two useful remarks. On the one hand, although the results of this work show that the MDI nuclear mean field with the incompressibility $K_0=230$ MeV can fairly reproduce the experimental data, systematic uncertainties such as cross sections and initialization, as well as cluster formation or light nuclei production, may all affect the quantitative observations of collective flows. Therefore, it would be interesting to explore how these factors affect the present quantitative results. On the other hand, our current results indicate that the momentum dependence of the mean field is correlated with the constraints on the nuclear EoS. Specifically, when the momentum dependence of the mean field is taken into account, the experimental data tend to favor a soft nuclear EoS, whereas the opposite conclusion is reached when this dependence is neglected. Therefore, it also deserves to further constrain the nuclear EoS more precisely using some other typical experimental observables, such as $K^+/K^0$. 
\section{Summary}\label{summary}
In summary, within an extended IBUU transport model, we have studied the proton and kaon as well as the associated $\Lambda$ production in Au + Au collisions at $\sqrt{s_{\rm NN}}=3$~GeV. Within three nuclear mean field scenarios, the proton transverse momentum spectra and rapidity dependent transverse momentum, the collective flows of both protons and kaon as well as the associated $\Lambda$ are systematically simulated and compared with the corresponding experimental data. It is found that only the MDI nuclear mean field with the incompressibility $K_0=230$ MeV could present a fair fit to all the experimental data,  while the MID nuclear mean field with either the soft EoS ($K_0=230$ MeV) or the stiff EoS ($K_0=380$ MeV) can only partially reproduce the experimental data. These findings indicate that the momentum dependence of nuclear mean field still plays a fundamental role in understanding the nuclear matter properties and describing experimental data of HICs even at high energies.

This work is supported by the National Natural Science Foundation of China under grant No.12475131, and the project of Guizhou Provincial Department of Science and Technology under Grant Nos. ZD[2026]108 and ZK[2023] General 248.

\end{document}